\documentclass[twocolumn,reprint,epsf,graphics,psfig,superscriptaddress]{revtex4}
\usepackage[bookmarks, colorlinks=true, plainpages = true,linkcolor = blue, citecolor = blue, urlcolor = blue, filecolor =blue]{hyperref}

\usepackage{amsfonts}
\usepackage{graphicx}
\usepackage{amsmath}
\usepackage{amssymb}
\usepackage{latexsym}
\usepackage{psfrag}
\usepackage{dcolumn}
\usepackage{datetime}
\usepackage{multirow}
\usepackage{subfigure}
\usepackage{amssymb,amsmath,rotate,color}

\begin{document}
\title{Mechanism of the enhanced conductance of a molecular junction under tensile stress}
\author{Alireza Saffarzadeh}
\altaffiliation{Author to whom correspondence should be addressed. Electronic mail: asaffarz@sfu.ca}
\affiliation{Department of Physics, Payame Noor University, P.O.
Box 19395-3697 Tehran, Iran} \affiliation{Department of Physics,
Simon Fraser University, Burnaby, British Columbia, Canada V5A
1S6}
\author{Firuz Demir}
\affiliation{Department of Physics, Simon Fraser University,
Burnaby, British Columbia, Canada V5A 1S6}
\author{George Kirczenow}
\affiliation{Department of Physics, Simon Fraser University,
Burnaby, British Columbia, Canada V5A 1S6}
\date{\today}

\begin{abstract}
Despite its fundamental importance for nano physics and chemistry
and potential device applications, the relationship between atomic
structure and electronic transport in molecular nanostructures is
not well understood. Thus the experimentally observed increase of
the conductance of some molecular nano junctions when they are
stretched continues to be counterintuitive and controversial. Here
we explore this phenomenon in propanedithiolate molecules bridging
gold electrodes by means of {\em ab initio} computations and
semi-empirical modeling. We show that in this system it is due to
changes in Au-S-C bond angles and strains in the gold electrodes,
rather than to the previously proposed mechanisms of Au-S bond
stretching and an associated energy shift of the highest occupied
molecular orbital and/or Au atomic chain formation. Our findings
indicate that conductance enhancement in response to the
application of tensile stress should be a generic property of
molecular junctions in which the molecule is thiol-bonded in a
similar way to gold electrodes.

\end{abstract}
\maketitle
\section{Introduction}

Molecular junctions in which a single molecule forms a stable
electrically conducting bridge between two metal electrodes have
been studied intensively experimentally and theoretically for more
than a decade due to their fundamental interest and potential
device applications \cite{Geo}. Achieving control over the
electronic transport properties of these nanostructures is crucial
for the realization of practical single-molecule electronic
devices. The electronic conductances of molecular junctions are
sensitive to their atomic geometries.\cite{Geo} These geometries
can be modified by varying the separation between the two metal
electrodes of the molecular junction, and this approach has been
used to modulate the conductances of a variety of molecular
junctions including those in which benzenedithiolate
(BDT),\cite{Bruot,Kim} octanedithiolate, \cite{Xu,Huang}
bipyridine\cite{Xu} or propanedithiolate (PDT) \cite{Hihath}
molecules bridge a pair of gold electrodes. In the case of
Au-BDT-Au,  conductance changes of more than three orders of
magnitude have been achieved in this way.\cite{Kim} Thus BDT has
been proposed recently as a candidate for applications as a broad
range coherent molecular conductor with tunable
conductance.\cite{Kim} However, despite its great importance for
molecular electronics, the present day understanding of the
relationship between the structure and electronic transport
characteristics of molecular junctions remains far from complete.
For example, intuitively, one might expect the conductance of a
molecular junction to decrease as the junction is stretched and
the interatomic bonds within it are weakened and therefore become
less conducting. However, a recent experiment \cite{Bruot} on
Au-BDT-Au junctions has demonstrated that, surprisingly, this need
not always be the case and that, to the contrary, stretching these
junctions can result in their conductances {\em increasing} by
more than an order of magnitude. This increase in the conductance
was attributed \cite{Bruot} to a strain-induced shift of the
highest occupied molecular orbital (HOMO) towards the Fermi level
of the electrodes resulting in resonant tunneling, as was
predicted by theories of the dependence of the HOMO energy on the
molecule-electrode separation, i.e., the Au-S bond length.
\cite{Andrews,Ke,Romaner,Li3,Toher,Pontes} However, another recent
study \cite{Kim} has indicated that the conductance of Au-BDT-Au
junctions is controlled mainly by the strength of the
molecule-electrode coupling rather than by the energy of a
molecular electronic level. Most recently it has been
proposed\cite{French} that gold atomic chains formed at the
molecule-electrode interfaces when the junction is stretched may
be responsible for the large increase observed\cite{Bruot} in the
conductance. Thus the origin of the observed behavior\cite{Bruot}
remains controversial. Interestingly,  the conductances of
Au-PDT-Au junctions have also been found experimentally to
increase as the junctions are stretched at cryogenic
temperatures\cite{Hihath}. While the bonding geometries between
the molecule and gold electrodes in Au-BDT-Au junctions have not
as yet been determined experimentally and remain the subject of
conjecture, \cite{Geo} experimental\cite{Hihath} and
theoretical\cite{Firuz1,Firuz2,Firuz3} inelastic tunneling
spectroscopy (IETS) studies have determined the experimentally
realized molecule-gold bonding geometries in Au-PDT-Au junctions.
For this reason Au-PDT-Au junctions may be more amenable as model
systems for developing a {\em definitive} understanding of the
counter-intuitive increase of the conductance as the junction is
stretched. We therefore explore this phenomenon in Au-PDT-Au
junctions theoretically in this Article.  We present a
comprehensive analysis of the relationship between the molecular
configuration, contact geometry, and molecular conductance, as the
junction is stretched. We find the calculated conductance to
increase by factors consistent with experiment,\cite{Hihath} as
the junction is stretched. Before the junction ruptures we find
the conductance to {\em decrease}, also as in the
experiment.\cite{Hihath} We find the conductance to decrease
shortly before and during the formation of a gold atomic chain at
a molecule-electrode interface, but that the conductance increase
resumes with further stretching once an atomic chain has formed.
However, we find the conductance increases for junctions with
atomic chains to be somewhat smaller than for those without them.
We show the increasing conductance as the junction is stretched to
be due mainly to increasing Au-S-C bond angles at the
molecule-electrode interfaces and strains in the electrodes
(geometrical changes that are expected to occur in many molecular
junctions under tensile stress)  rather than to atomic chain
formation or elongation of the Au-S bonds. We find the latter to
be too small to affect the measured conductance\cite{Hihath}
strongly.

\section{Estimation of the Junction Geometries}

\begin{figure}[t!]
\centerline{\includegraphics[width=0.85\linewidth]{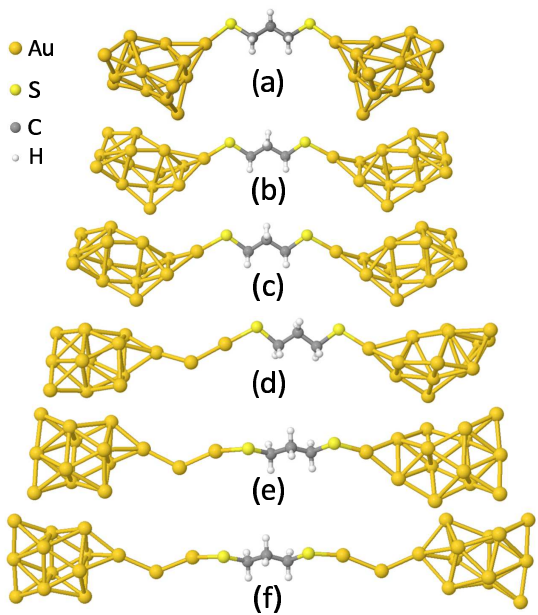}}
\caption{(Color online) Relaxed structures of the Au-PDT-Au
molecular junction for various distances $d$ between the outermost atoms
of the Au clusters. (a) $d$=21.75{\AA}, (b) $d$=25.70{\AA}, (c)
$d$=26.50 {\AA}, (d) $d$=29.00 {\AA}, (e) $d$=30.00 {\AA},
(f) $d$=33.40 {\AA}. As the junction is stretched Au-Au bonds near
the molecule break in succession until in (f) the molecule is connected
to each Au cluster by a chain of Au atoms.}
\label{Figure1}
\end{figure}

We study the evolution of the molecular junction's geometry as the
Au-PDT-Au junction is stretched by carrying out \textit{ab initio}
calculations of the relaxed geometries of extended molecules each
consisting of a PDT molecule and two 14-atom gold clusters bound
to the PDT sulfur atoms as shown in Fig. \ref{Figure1}. Our
previous work \cite{Firuz1,Firuz2,Firuz3} showed the electronic
transport properties of Au-PDT-Au extended molecules to have
converged sufficiently with increasing gold cluster size that
14-atom gold clusters can serve as a realistic model of the
macroscopic gold electrodes that contact the PDT molecule in
experiments.\cite{Hihath} Our starting point is the relaxed
geometry shown in Fig. \ref{Figure1}(a) that our IETS studies
\cite{Firuz1,Firuz2,Firuz3} identified as having the gold-sulfur
bonding geometry that was most commonly realized in experiments
\cite{Hihath} on Au-PDT-Au junctions. The other structures that we
considered were generated from this one by moving the atom of each
gold cluster that was furthest from the molecule slightly further
from or closer to the molecule and then freezing the positions of
these two outer atoms and relaxing the positions of all of the
other atoms of the extended molecule. This procedure was iterated
to produce relaxed junction geometries with differing distances
between the two outer Au atoms. Examples are shown in Fig.
\ref{Figure1}. The relaxations were all carried out using the
GAUSSIAN 09 package with the B3PW91 functional and Lanl2DZ
pseudopotentials and basis sets \cite{gaussian,Pedrew} to locate
local minima of the total energy of the extended molecule computed
within density functional theory.

Since our approach to calculating the molecular junction
geometries is fully quantum mechanical and the total energies of
the trial geometries sampled in the relaxation process are
computed within density functional theory, our relaxed geometries
are expected to be more accurate than those obtained using
computationally less demanding methods that employ parameterized
interatomic potentials and classical mechanics, such as those in
Ref. \onlinecite{French}.

Since our geometry relaxations were performed effectively at zero
temperature, the relaxed structures that we obtained may be
expected to be representative of those realized in the cryogenic
temperature experiments of Hihath {\em et al.} \cite{Hihath} on
Au-PDT-Au junctions, especially in view of the fact that our
starting molecule-electrode bonding geometry before the junction
was stretched  was that previously identified as the one most
commonly realized in those experiments.\cite{Firuz1,Firuz2,Firuz3}
The fact that the bonding geometry most commonly realized in the
experiments on Au-PDT-Au junctions\cite{Hihath} is
known\cite{Firuz1,Firuz2,Firuz3} obviated the need for us to
perform relaxations/stretching for a large ensemble of different
starting junction geometries such as was necessary in the study in
Ref. \onlinecite{French} of Au-BDT-Au junctions for which the
bonding geometry most commonly realized in experiments has yet to
be determined.

\section{Conductance Calculations}

We calculate the zero bias conductance $g$ at the Fermi energy for
the Au-PDT-Au molecular wire at various elongations from the
Landauer formula \cite{Geo}
\begin{equation}\label{g}
g=g_0\sum_{i,j}|t_{ji}|^2\frac{v_j}{v_i}
\end{equation}
where $g_0=2e^2/h$, $t_{ji}$ is the transmission amplitude through
the PDT molecule, $i$ is the electronic state of a carrier with
velocity $v_i$ that is coming from the left lead, and $j$ is the
electronic state of a carrier with velocity $v_j$ that has been
transmitted to the right lead. To couple the extended molecule to
the electron reservoirs, as in previous work
\cite{Firuz1,Firuz2,Firuz3,Cardamone,Piva1,Piva2,Dalgleish1,George2,Renani,Saffar},
we attach a large number of semi-infinite quasi-one-dimensional
ideal leads to the valence orbitals of the outer gold atoms of the
extended molecule. We find the transmission amplitudes $t_{ji}$ by
solving the Lippmann-Schwinger equation
\begin{equation}\label{1}
|\Psi^\alpha\rangle=|\Phi_0^\alpha\rangle+G_0(E)W|\Psi^\alpha\rangle
\end{equation}
where $|\Phi_0^\alpha\rangle$ is an electron eigenstate of the
$\alpha$th ideal semi-infinite one-dimensional left lead that is
decoupled from the extended molecule, $G_0(E)$ is the Green's
function of the decoupled system of the ideal leads and the
extended molecule, $W$ is the coupling between the extended
molecule and the ideal leads, and $|\Psi^\alpha\rangle$ is the
scattering eigenstate of the complete coupled system associated
with the incident electron state $|\Phi_0^\alpha\rangle$. The
semi-empirical  extended H\"{u}ckel model \cite{Geo} with the
parameters of Ammeter {\em et al.} \cite{yaehmop} was used to
evaluate the Hamiltonian matrix elements and atomic valence
orbital overlaps that enter the Green's function $G_0(E)$ in Eq.
(\ref{1}). As has been discussed in Refs. \onlinecite{Geo} and
\onlinecite{Cardamone}, this methodology involves no fitting to
any experimental data relating to transport in molecular
junctions. It is known to yield low bias conductances in
reasonably good agreement with experiments for PDT bridging gold
electrodes\cite{Firuz1,Firuz2} as well as other molecules
thiol-bonded to gold electrodes
\cite{Kushmerick02,Cardamone,Geo,Datta1997,
EmberlyKirczenow01,EmberlyKirczenow01PRB}.

We note that density functional theory (DFT)-based transport
calculations are often used to calculate the conductances of
molecular junctions.\cite{Geo} For example, transport calculations
based on DFT with approximate self-interaction corrections have
been used in the recent theoretical study\cite{French} of the
conductance enhancement of Au-BDT-Au junctions under tensile
stress. However, density functional theory is a tool designed
specifically for calculating ground state total
energies.\cite{HK,KS} Thus it is appropriate to use DFT to
calculate molecular junction geometries, as is done in the present
work, since these geometries are found by minimizing the total
energy. On the other hand, transport is {\em not} a ground state
property. For this reason DFT-based transport calculations are not
well justified at the fundamental level and the results obtained
from them in practice are often unreliable.\cite{Geo} Attempts are
being made to correct for these limitations of DFT,\cite{Geo}
however, these corrections involve uncontrolled approximations and
their effectiveness is uncertain at this time. Moreover, DFT-based
transport calculations are especially problematic for the case in
which the distances between the molecule and the contacts in a
molecular junction are being varied, as has been discussed by
Koentopp {\em et al.}\cite{Koentopp} and others.\cite{Geo} For
these reasons it is unclear whether any counterintuitive transport
effect predicted by a DFT-based theory in a molecular junction
under tensile strain is physical or an artifact of the
shortcomings of DFT and/or any corrections that have been applied
to it. We therefore prefer to use instead the semi-empirical
extended H\"{u}ckel theory-based approach to transport that has
been outlined and justified in the preceding paragraph.

\section{Results}

\begin{figure}[t!]
\centerline{\includegraphics[width=0.9\linewidth]{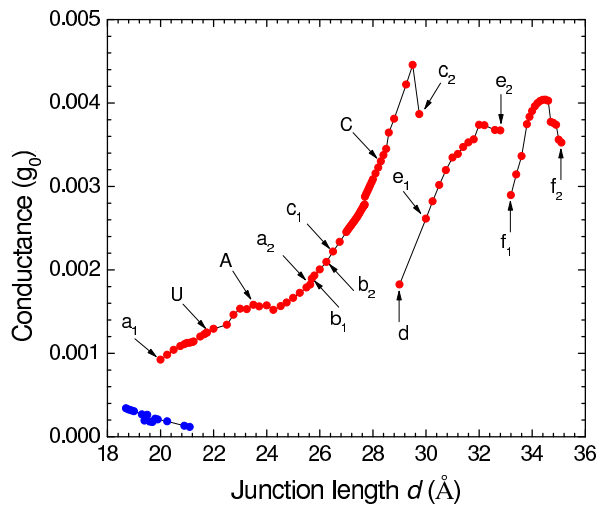}}
\caption{(Color online)  Calculated conductance at zero
temperature of trans (gauche) Au-PDT-Au junctions vs junction
length for different structures is shown in red (blue). Red
circles between a$_1$ and a$_2$, b$_1$ and b$_2$, c$_1$ and c$_2$,
e$_1$ and e$_2$, and between f$_1$ and f$_2$ show the conductances
for structures similar to Fig. 1(a), Fig. 1(b), Fig. 1(c), Fig.
1(e), Fig. 1(f), respectively. Point $d$ indicates the structure
shown in Fig. 1(d). A and C are explained in the inset of Fig. 3.
U is the unstrained junction.} \label{Figure2}
\end{figure}

The calculated conductances of our relaxed structures are plotted
in Fig. \ref{Figure2} vs. the junction length $d$, i.e., the
distance $d$ between the outermost atoms of the two gold clusters.
The structure in Fig. \ref{Figure1} (a) (with $d=21.75$ {\AA} and
labeled U in Fig. \ref{Figure2}) is an unstrained junction since
it was relaxed without constraints on any atoms.

Fig. \ref{Figure2} shows that when a compressive strain is applied
to the unstrained junction, the conductance of the junction
decreases (red circles to the left of U in Fig. \ref{Figure2}). A
larger drop in the conductance occurs when the molecular geometry
switches from trans (red circles) to gauche (blue circles) under
compressive strain. Lower conductances for gauche than trans
conformations of other molecular wires have been predicted
previously.\cite{Li2,Paulsson2009} Note that, in the present work
gauche conformations (blue circles) were formed only for $d \le
21.10$ {\AA}; for all other junction lengths only trans molecular
conformations were found.

On the other hand, stretching the junction from $d$=21.75 {\AA} (U
in Fig. \ref{Figure2}) to $d$=29.50 {\AA} results in an increasing
junction conductance followed by a decrease at $d$=29.75 {\AA}
(labeled c$_2$ in Fig. \ref{Figure2}) which signals that the
junction is close to rupturing or forming an Au atomic chain as in
Fig. \ref{Figure1} (e). In the junction length interval from
$d$=20.00 {\AA} to $d$=25.64 {\AA}, i.e. from a$_1$ to a$_2$ in
Fig. \ref{Figure2}, the nearest gold atoms to the PDT molecule
bond to three neighboring gold atoms and the junction geometry for
the trans molecular conformation stays qualitatively similar to
that in Fig. \ref{Figure1}(a). At $d$=25.70 {\AA}, labeled b$_1$
in Fig. \ref{Figure2}, a bond breaks between the Au atom at the
tip of the right gold cluster and one of its neighbors, as shown
in Fig. \ref{Figure1}(b). However, interestingly, this bond
breaking does not result in any drop in the conductance which, to
the contrary, continues to rise. The nature of the resulting
bonding geometry is maintained until $d$=26.25 {\AA}, labeled
b$_2$. At $d=26.50$ {\AA} labeled c$_1$ a bond breaks between the
Au atom at the tip of the left cluster and one of its neighbors,
resulting in a structure similar to that in Fig. \ref{Figure1}(c).
However, the conductance continues to rise smoothly as the
junction is stretched although with an increasing slope until
$d=29.75$ {\AA} (c$_2$) where the conductance declines although
the structure remains qualitatively similar to that in Fig.
\ref{Figure1}(c).

We find further stretching to result in either rupture of the
junction or formation of a chain of gold atoms between the
molecule and one of the gold clusters, depending on the details of
the stretching procedure. Formation of the atomic chain results in
the structure shown in Fig. \ref{Figure1}(e) and a significant
drop in the conductance to the value labeled e$_1$ in Fig.
\ref{Figure2}. This conductance drop is reasonable because gold
atomic chains have only a single conducting quantum channel as
compared to the multiple channels in the gold clusters.
Compressing the structure in Fig. \ref{Figure1}(e) results in the
structure in Fig. \ref{Figure1}(d) and a decline in the junction
conductance from e$_1$ to d in Fig. \ref{Figure2}. Interestingly,
although the structure in Fig. \ref{Figure1}(d) involves a gold
atomic chain, its junction length $d=29.00$ {\AA} is smaller than
that of c$_2$ ($d=29.75$ {\AA}) although the latter structure has
no atomic chain and resembles Fig. \ref{Figure1}(c). If the
junction in Fig. \ref{Figure1} (d) or (e) is stretched, the
conductance resumes its increase until the junction length
approaches $d=32.80$ at e$_2$ in Fig. \ref{Figure2} where a small
decrease in the conductance again heralds the appearance of a Au
atomic chain, this time at the right Au cluster (Fig.
\ref{Figure1} (f)). The formation of this chain with $d=33.00$
{\AA} again results in an abrupt drop in the conductance to the
value labeled f$_1$ in Fig. \ref{Figure2}. Thus an abrupt drop in
conductance when an Au atomic chain is formed appears to be a
generic property of this system. Upon further stretching of the
junction from $d=33.00$ {\AA} to $d=35.10$ {\AA}, i.e. from f$_1$
to f$_2$, the conductance increases and then declines while the
number of Au-Au bonds between the tips of the gold clusters and
the adjacent gold atoms remains the same; the elongation of the
junction results in changes in various bond angles in the system.

Our numerical results show that the responses of the electronic
energy eigenvalues and eigenvectors of the extended molecule to
stretching of the junction both contribute to the rises in the
conductance. The details are complex since several pairs of almost
degenerate eigenstates mediate conduction through the molecule and
the contributions of the two states of each degenerate pair
interfere destructively with each other. The reason for the
destructive interference is that the two degenerate states have
opposite parity (within the PDT molecule) with respect to the
approximate left-right mirror symmetry of the PDT molecules in
Fig. \ref{Figure1}.

\begin{figure}[t!]
\centerline{\includegraphics[width=0.9\linewidth]{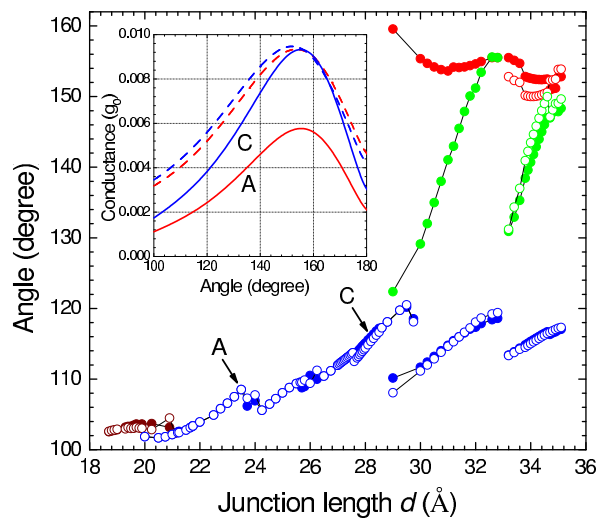}}
\caption{(Color online) Variation of the C-S-Au (blue), S-Au-Au
(red) and Au-Au-Au (green) angles for trans PDT junctions with
junction length $d$. C-S-Au angles for gauche structures are
maroon. Solid (hollow) circles show angles to the left (right) of
the molecule. The solid (dashed) lines in the inset show the
conductance vs. C-S-Au angle for junctions obtained from those
marked A and C in Fig. \ref{Figure2} and \ref{Figure3} by rigidly
rotating the 14 (1)-atom gold clusters about the S atoms.}
\label{Figure3}
\end{figure}

However, we shall show that the physics underlying conductance
increases as the junction is stretched can be identified by
examining the relationship between the conductance and the
geometrical changes that occur in the junction. Stretching results
in changes in the chemical bond lengths and bond angles in the
junction. For example, when the junction length increases from
$d=29.00$ {\AA} to 32.80 {\AA} (i.e., from d to e$_2$ in Fig.
\ref{Figure2}), the S-Au bond lengths increase by only a small
amount (0.05 {\AA}), while the C-S-Au bond angles at the left
(right) end of the molecule change from 110.15$^\circ$ to
118.60$^\circ$ (108.07$^\circ$ to 119.40$^\circ$), indicating that
most of the elongation is due to changes in bond angles.
Importantly, during this elongation the conductance of the
junction increases by a factor of $\sim 2$ but we find that only a
small $\sim 16\%$ conductance change can be attributed to the
elongation of the Au-S bonds. We find that much greater elongation
of the Au-S bonds (by 0.6 {\AA}) would result in large conductance
increases (by factors $\sim 4$) due to a transport resonance
approaching the electrode Fermi level, qualitatively similar to
the findings of previous theories relating the conductance of
Au-BDT-Au junctions to the molecule-electrode separation.
\cite{Andrews,Ke,Romaner,Li3,Toher,Pontes} However, our {\em ab
initio} junction geometry calculations show such large Au-S bond
elongations to be an order of magnitude larger than those
predicted to be realized during the experimental conductance
measurements \cite{Hihath}, and therefore not relevant to the
observed conductance increases.

Thus we turn now to the role of the bond angles in the
conductance. Fig. \ref{Figure3} shows the dependence on the
junction length $d$ of the C-S-Au angles, and also the S-Au-Au and
Au-Au-Au angles when an Au atomic chain forms. The corresponding
angles at the left and right ends of the molecule (solid and
hollow circles) are very similar. The dependence of the C-S-Au
bond angles (blue circles) in Fig. \ref{Figure3} on the junction
length $d$ is strikingly similar to that of the conductance in
Fig. \ref{Figure2}; the main features of either plot are
reproduced in the other almost quantitatively. In contrast,
although the S-Au-Au and Au-Au-Au angles depend on $d$ and the
latter angle even changes by more than 30$^\circ$ in the
stretching process, the role of these angles is not crucial in the
conductance increase, since conduction through Au atoms is mainly
via the isotropic $s$ orbitals.

To better understand the role of the bond angles, we chose two
junctions marked A and C in Fig. \ref{Figure2} and \ref{Figure3}.
For these we computed the dependence of the conductance on the
C-S-Au angles {\em alone}, varying these angles by rotating the
gold clusters rigidly about the S atoms, holding the PDT molecule
fixed. This yielded the solid curves in the inset of Fig.
\ref{Figure3}. These plots show that for each structure A or C the
dependence of the conductance on the C-S-Au angles accounts for
roughly half of the conductance change between a$_1$ and c$_2$ in
Fig. \ref{Figure2},  the remainder being due mainly to the
structural differences between the A and C gold clusters. A
similar calculation but including only one Au atom from each Au
cluster yielded the dashed curves in the inset of Fig.
\ref{Figure3}. Since the difference between the dashed red and
blue curves is small and is due to the stretching of the Au-S
bonds and structural changes within the PDT molecule itself in
going between structures A and C, we conclude that these effects
make at most minor contributions to the conductance increase when
the junction is stretched.

As has been discussed above, we find the conductances of Au-PDT-Au
junctions to increase by similar amounts as the junction is
stretched, whether a gold atomic chain is present at the interface
between the molecule and one of the gold electrodes, or at the
interfaces between the molecule and both of the gold electrodes,
or at neither interface. However, the maximal value of this
conductance increase is somewhat larger if there is no gold atomic
chain present than in either of the latter two cases. By contrast
in Ref. \onlinecite{French} it has been predicted that for
Au-BDT-Au junctions the conductance increase in response to
stretching of the junction is much larger if a gold atomic chain
forms than if it does not. The mechanism proposed in Ref.
\onlinecite{French} for this very large conductance increase is
unrelated to the mechanism of conductance enhancement in Au-PDT-Au
junctions that we propose in the present Article. The conductance
increase predicted in Ref. \onlinecite{French} is due to a strong
feature in the Au $s$ and $p$ density of states of the atomic
chains near the Fermi energy that results in enhanced electron
transmission through the molecular junction. Some caution
regarding this prediction is in order since it is the result of
DFT-based transport calculations although a self-interaction
correction intended to address some of the deficiencies of DFT was
included in the calculation.\cite{French}

\section{Conclusions}

In summary, we have presented a systematic exploration of the
response of the conductance of propanedithiolate molecular
nanowires bridging gold electrodes to mechanical stretching of
this junction, based on {\em ab initio} density functional theory
and semi-empirical techniques. Our results demonstrate
theoretically that the counterintuitive phenomenon of increasing
conductance in response to junction elongation should occur in a
gold-alkanedithiolate-gold junction. We showed that three
different junction geometries can each show a conductance increase
in response to junction elongation followed by a conductance
decrease and either rupture or a transition to a different
junction geometry. A very similar conductance increase followed by
a decrease and subsequent junction rupture has been observed
experimentally in an Au-PDT-Au junction and can be seen in Fig. 4a
of Ref. \onlinecite{Hihath}. We showed the conductance increases
to be due primarily to increases in the C-S-Au bond angles at the
molecule-electrode interfaces and to atomic rearrangements in the
gold electrodes. Conductance increases in response to increasing
C-S-Au bond angles are reasonable because of the important role
that carbon and sulfur atomic valence $p$-orbitals play in
transport through organic molecules and the anisotropic nature of
the overlaps between these orbitals on different atoms. For this
reason we expect the mechanism of the conductance increase under
tensile stress that we have identified to be active in all
molecular junctions with molecules thiol bonded in a similar way
to gold electrodes. Therefore we predict that this interesting
phenomenon is a universal generic property of such junctions. We
have also demonstrated that changes in the Au-S bond length and in
the structure of the propanedithiolate molecule that occur in
response to elongation of the junction play a minor role in the
observed conductance increase. Our findings also demonstrate that
electron transport through the PDT molecule can be controlled by
varying the length of the junction and that both continuous
modulation of the conductance and abrupt switching between high
and low conductance states can be induced in this way. These
remarkable changes in the molecular conductance indicate that
alkanedithiol molecular nanowires have potential applications as
mechanical switches in nanoscale devices.

This work was supported by NSERC, CIFAR, WestGrid, and Compute
Canada.

\end{document}